\documentclass{aa}  

\usepackage{graphicx}
\usepackage{txfonts}
\usepackage{xcolor}
\usepackage[colorlinks=true, allcolors=blue]{hyperref}

\usepackage{natbib}
\bibpunct{(}{)}{;}{a}{}{,}

\newcommand{\ergs}[1]{$\times 10^{#1}$ erg s$^{-1}$}
\newcommand{\oergs}[1]{$10^{#1}$ erg s$^{-1}$}
\newcommand{\uergs}{erg s$^{-1}$}
\newcommand{\hcm}[1]{$\times 10^{#1}$ cm$^{-2}$}
\newcommand{\ohcm}[1]{$10^{#1}$ cm$^{-2}$}

\newcommand{\nh}{$N_{\rm H}$\xspace}
\newcommand{\lx}{\hbox{L$_{\rm x}$}}

\newcommand{\cts}{cts s$^{-1}$\xspace}

\newcommand{\Hone}{\ion{H}{i}\xspace}

\newcommand{\ltsima}{$\buildrel < \over \sim$}
\newcommand{\lsim}{\lower.5ex\hbox{\ltsima}}

\newcommand{\rahour}{\hbox{\ensuremath{^{\rm h}}}}
\newcommand{\ramin}{\hbox{\ensuremath{^{\rm m}}}}

\newcommand{\xspec}{\texttt{XSPEC}\xspace}
\newcommand{\eSASS}{\texttt{eSASS}\xspace}
\newcommand{\decam}{{DECam}\xspace}
\newcommand{\gaia}{{\it Gaia}\xspace}

\newcommand{\einstein}{{\it Einstein}\xspace}
\newcommand{\xmm}{{\it XMM-Newton}\xspace}

\newcommand{\rosat}{ {\it ROSAT}\xspace}

\newcommand{\nicer}{{\it NICER}\xspace}
\newcommand{\srg}{{\it SRG}\xspace}
\newcommand{\ero}{\mbox{eROSITA}\xspace}
\newcommand{\esrc}{\mbox{eRASSU\,J060839.5$-$704014}\xspace}
\newcommand{\xsrc}{\mbox{3XMM\,J051034.6$-$682640}\xspace}
\newcommand{\CnC}{\mbox{HM\,Cnc}\xspace}
\newcommand{\Vul}{\mbox{V407\,Vul}\xspace}
\graphicspath{{figs/}}

\begin{document} 

\title{\esrc: A double degenerate ultra-compact binary in the direction of the LMC\thanks{Based on observations obtained with \xmm, an ESA science mission with instruments and contributions directly funded by ESA Member States and NASA}}

\author{ C. Maitra\inst{\ref{mpe}} \and
        F. Haberl\inst{\ref{mpe}} \and
        G. Vasilopoulos\inst{\ref{oas},\ref{uoa}}
        A. Rau\inst{\ref{mpe}} \and
        A. Schwope\inst{\ref{potsdam}} \and
        S. Friedrich\inst{\ref{mpe}} \and
         D.A.H. Buckley\inst{\ref{salt},\ref{uct}} \and
        F. Valdes\inst{\ref{noir}} \and
        D. Lang\inst{\ref{pitp}} 
       \and S. A. Macfarlane\inst{\ref{salt}}
       } 

\titlerunning{\esrc A double degenerate ultra-compact binary in the LMC}
\authorrunning{Maitra et al.}

\institute{
Max-Planck-Institut f{\"u}r extraterrestrische Physik, Gie{\ss}enbachstra{\ss}e 1, D-85748 Garching, Germany\label{mpe}, \email{cmaitra@mpe.mpg.de}
\and
Universit\'e de Strasbourg, CNRS, Observatoire astronomique de Strasbourg, UMR 7550, F-67000 Strasbourg, France\label{oas}
\and
Department of Physics, National and Kapodistrian University of Athens, University Campus Zografos, GR 15783, Athens, Greece\label{uoa} 
\and
Leibniz-Institut f{\"u}r Astrophysik Potsdam (AIP), An der Sternwarte 16, 14482 Potsdam, Germany\label{potsdam}
\and
South African Astronomical Observatory, PO Box 9, Observatory Rd, Observatory 7935, South Africa\label{salt}
\and
Department of Astronomy, University of Cape Town, Private Bag X3, Rondebosch 7701, South Africa\label{uct}
\and
NSF’s National Optical/Infrared Research Laboratory (NOIRLab), 950 N. Cherry Ave, Tucson, AZ 85732, USA\label{noir}
\and
Perimeter Institute for Theoretical Physics, 31 Caroline Street North, Waterloo, ON  N2L 2Y5, Canada\label{pitp}
}

\date{Received 2023 / Accepted 2023}

\abstract
   {During four all-sky surveys (eRASS1--4), \ero, the soft X-ray instrument aboard {\it Spektrum-Roentgen-Gamma (SRG)} detected a new supersoft X-ray source, \esrc, in the direction of the Large Magellanic Cloud (LMC).} 
   {We arranged follow-up observations in the X-ray and optical wavelengths and further searched in archival observations to reveal the nature of the object.}
   {Using X-ray observations with \xmm we investigated the temporal and spectral behaviour of the source.}
   {We discover pulsations at ~374\,s with a pulse profile consistent with 100\% modulation. We identify two other periodicities in the \ero data, which we establish as aliases due to the sampling of the \ero light curve.
   We identify a multi-wavelength counterpart to the X-ray source in UVW1 and g, r, i, and z images obtained by the optical/UV monitor on \xmm 
   and the Dark Energy Camera at the Cerro Tololo Inter-American Observatory. 
   The timing and spectral characteristics of the source are consistent with a double degenerate ultra-compact binary system in the foreground of the LMC. \esrc\ belongs to a
rare class of AM CVns, which are important to study in the context of progenitors of SN\,Ia and for persistent gravitational wave detection. }
{We identify \esrc as a new double degenerate ultra-compact binary located in the foreground of the LMC.}


\maketitle   


\section{Introduction}
\label{sec:intro}

Double degenerate ultra-compact binary systems comprise two compact objects (white dwarfs, neutron stars, or black holes) at the post-common envelope phase of binary evolution. Double degenerate interacting white dwarf systems (DDs from now) are particularly interesting from the cosmological perspective as possible progenitors of SNe Ia \citep[][]{2018MNRAS.473.5352L}. In addition, these systems, which are expected to be in tight binary orbits, can merge due to the loss of orbital angular momentum and are expected to be strong emitters of gravitational waves (GWs) detectable by future space-based gravitational wave detectors like, for example, the Laser Interferometer Space Antenna (LISA) \citep[][]{2020ApJ...893....2L}. For a comprehensive review of double degenerate systems including DDs, see \citet[][and references within]{2014LRR....17....3P}.

DDs have been identified in X-rays as a sub-class of heterogenous objects classified as supersoft X-ray sources (SSSs), first identified with \rosat \citep[][]{1991Natur.349..579T}. SSSs are characterised by very soft X-ray spectra with kT $\sim$ 15--80 eV \citep[e.g.][]{1997ARA&A..35...69K} and a 
wide range of luminosities. The most luminous ($\sim$10$^{36}$ to $\sim$\oergs{38}) can be explained by stable nuclear burning white dwarfs (WDs), which in most cases accrete H-rich matter from a companion star \citep{1992A&A...262...97V}. Other objects in this class include WDs as central stars of planetary nebulae (PNe) \citep[][]{2008A&A...482..237K,2010A&A...519A..42M}, and magnetic cataclysmic variables (mCVs), including polars and soft intermediate polars \citep[see e.g.,][]{1990SSRv...54..195C,1995A&A...297L..37H,1996A&A...310L..25B,2008A&A...489.1243A,2022ApJ...932...45O}. 
A majority of these systems were discovered in the direction of the Magellanic Clouds, which are frequently observed by X-ray observatories. The low Galactic foreground absorption in their direction makes them ideal laboratories for the detection and investigation of SSSs \citep[]{2022A&A...657A..26M}.

Three sources, RX\,J0806.3$+$1527 \citep[\CnC,][]{1999A&A...347...47B}, its twin RX\,J1914.4$+$2456 \citep[\Vul,][]{1995A&A...297L..37H}, and \xsrc \citep[][]{2017A&A...598A..69H}, discovered as SSSs in \rosat and \xmm data, are considered classical examples of DDs with periodicities of 5.4\,min, 9.5\,min and 23.6\,min, respectively. These periods were also found in the light of their optical companions (\CnC: \citet{2002MNRAS.332L...7R}, \citet{2002A&A...386L..13I}, \Vul: \citet{2000MNRAS.311...75R}, \xsrc: \citet{2018A&A...617A..88R}). 

The X-ray flux drops to zero between pulses, and no other periods are seen \citep[][]{2014A&A...561A.117E,1998MNRAS.293L..57C,2017A&A...598A..69H}. 

Several models have been proposed to explain the X-ray emission from these systems. The two main accretion models in this regard include mass transfer from a Roche-lobe-filling WD to either a magnetic (polar-like) or a non-magnetic (Algol-like) accretor. In the polar-like model \citep[][]{1998MNRAS.293L..57C}, the magnetic field of the accreting WD inhibits the formation of an accretion disc and matter reaches the magnetic polar cap. In the Algol-like type, also known as a “direct impact” accretion model \citep[][]{2001A&A...368..939N,2002MNRAS.331L...7M}, a light companion is assumed so that a disc would not form, resulting in the stream directly hitting the surface of the accreting WD. However, models invoking accretion predict an orbital widening for the two degenerate WDs, in contrast to what is observed in RX\,J0806.3$+$1527 and RX\,J1914.4$+$2456, although solutions to circumvent this issue have been proposed \citep[see][and references therein]{2012ApJ...758...64K}. The main alternative to the accretor model is the “unipolar inductor” model \citep[e.g.][]{2006A&A...447..785D,Colpi2009PhysicsOR}. This model involves a magnetic primary WD and a (non-magnetic) secondary that does not fill its Roche lobe. In this case, if the spin period of the primary and the orbital period are not synchronous, then the secondary crosses the primary’s magnetic field as it moves along the orbit. The resulting electromotive force drives an electric current between the two WDs (assuming the presence of ionised material between them), whose dissipation heats the polar caps on the primary. This method has however been highlighted as not being efficient enough to explain the observed X-ray flux \citep[][]{2012ApJ...757L...3L}. 

Nonetheless, the 
emission mechanism for the population of DDs remains an open question and more observable systems are required to fine-tune the models. 

Here, we report on the discovery of a new SSS in the direction of the Large Magellanic Cloud (LMC), \esrc, which we identify as a DD candidate in the foreground. In Sect.~\ref{sec:xobs} we present the X-ray observations of the object, including \ero and \xmm observations. Section~\ref{sec:xanalysis} describes the temporal and spectral analysis of the X-ray data. Section~\ref{sec:optical} describes the identification of the optical counterpart provided from Dark Energy Camera (\decam) archival exposures and GROND observations, and Sect.~\ref{sec:discussion} presents the discussion and our conclusions.

\section{X-ray observations}
\label{sec:xobs}

\subsection{\ero}
\label{sec:eroobs}

\esrc was discovered as a bright new source by \ero \citep{2021A&A...647A...1P}, the soft X-ray instrument on board the {\it Spektrum-Roentgen-Gamma} (\srg) mission \citep{2021A&A...656A.132S}, which surveyed the X-ray sky between December 2019 and February 2022 in the energy range of 0.2--8\,keV.
The source was detected in all four all-sky surveys (eRASS1 to eRASS4) and was scanned a total of 375 times between 12:57 UTC on March 13 2020 (MJD  58921.53976) and 22:02 UTC on September 27 2021 (MJD 59484.918664). During this interval, it collected a total exposure of 5.2\,ks (after taking into account vignetting and dead-time corrections); see Table\,\ref{tabobsero}. 
For the data analysis, we used the \ero Standard Analysis Software System 
\citep[\eSASS version {\tt eSASSusers\_211214\_0\_3};][]{2022A&A...661A...1B}. 
To extract source and background events corresponding to light curves and spectra, we used the \eSASS task \texttt{srctool}  
\citep[see e.g. ][]{2021A&A...647A...8M,2022A&A...661A..25H}. 
For the source products, we selected all valid pixel patterns (PATTERN=15) and used circular regions with radii of 50\arcsec\ and 75\arcsec\ around the position of the source and a nearby source-free region. For the light curves, we combined the data from all cameras (telescope modules (TMs) 1--7) and applied a cut in the fractional exposure of 0.15 (FRACEXP$>$0.15). The fractional exposure corresponds to the product of the fractional collecting area and the fractional temporal coverage that overlaps with the time bin.
 We created a combined spectrum from the data of TMs 1--4 and 6, the five cameras with an on-chip optical blocking filter. TM5 and TM7 suffer from a light leak \citep{2021A&A...647A...1P} and no reliable energy calibration is available yet. 
We also performed barycentre corrections by converting the arrival time of the photons from the
local satellite into the Solar System barycentric frame using the \texttt{HEASOFT} task \texttt{barycen}, the JPL-DE405 ephemeris table, and the target coordinates. We refer to \citet[][]{2022A&A...661A..41S} for details. Finer timing corrections such as clock drifts and frame time jitters would be considered in the next data processing version but do not affect the precision of a relatively slow periodic signal like that detected in \esrc.

\subsection{\xmm}
\label{sec:xmmobs}
\begin{figure}
\centering
\resizebox{0.6\hsize}{!}{\includegraphics{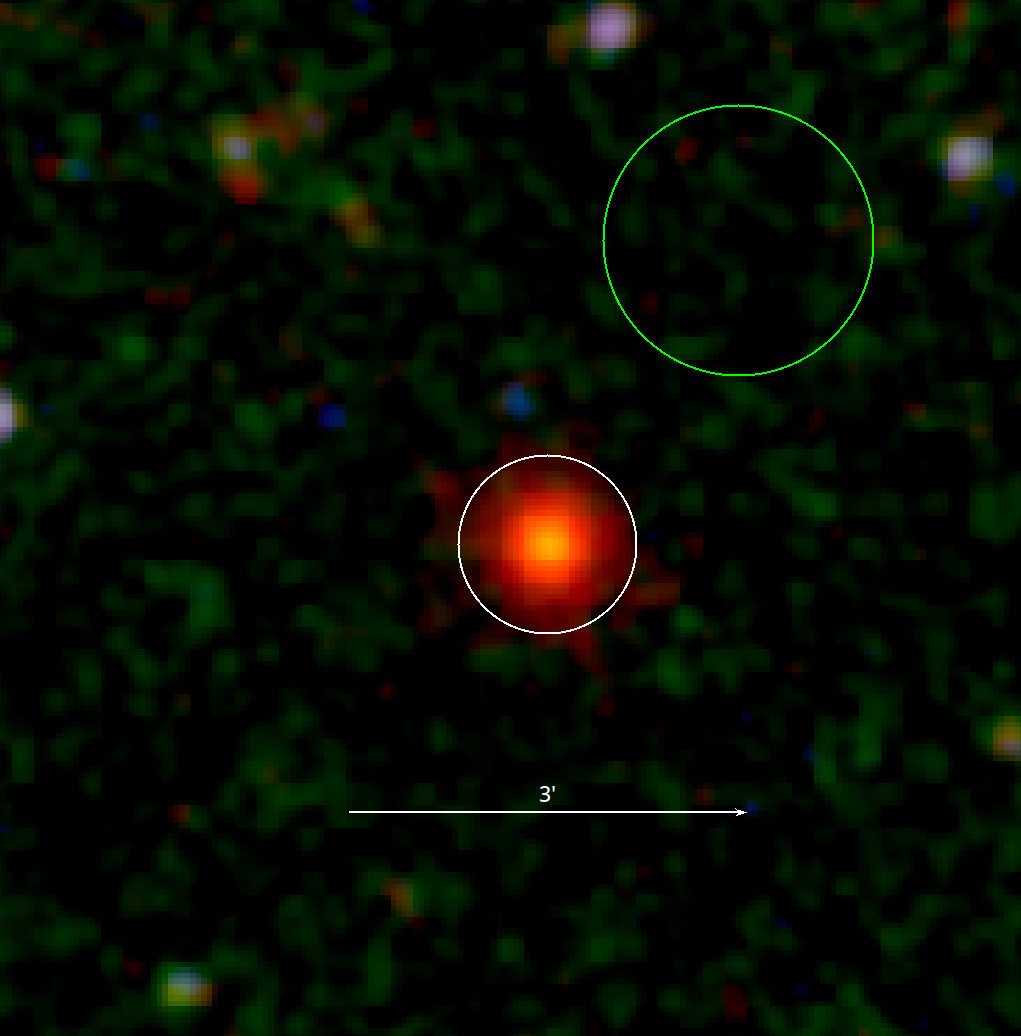}}
  \caption{Region around \esrc extracted from the full RGB mosaic of the EPIC LMC observations showing the source (white) and background extraction circles (green).}
  \label{fig:mosaic}
\end{figure}

To investigate the nature of \esrc in detail, we triggered one of our \xmm anticipated target of opportunity observations (PI Maitra) to follow up on new supersoft sources in the Magellanic system. 
\xmm consists of the European Photon Imaging Camera (EPIC, 0.15--12\,keV band), with two of the three \xmm telescopes equipped with metal oxide semi-conductor (MOS) CCD arrays \citep{2001A&A...365L..27T} and the third with a pn-CCD \citep[][]{2001A&A...365L..18S}.
The observation was performed on January 22 2022 (observation ID 0882050401, start MJD = 59600.6412). 
We used the EPIC cameras with thin optical blocking filters, in full-frame readout mode.
\xmm/EPIC data were processed using the \xmm data analysis software SAS, version 19.1.0\footnote{Science Analysis Software (SAS):\\ \url{https://www.cosmos.esa.int/web/xmm-newton/sas}}. For the best X-ray source position, we used the one derived by the \xmm pipeline of 
$\alpha_\mathrm{J2000.0}= 06\rahour\,08\ramin\,38\fs98$ and $\delta_\mathrm{J2000.0} = -70\degr\,40\arcmin\,13\farcs2$ with a 1\,$\sigma$ statistical uncertainty of $0\farcs1$ and a remaining systematic error of $0\farcs5$ after astrometric correction. The mean EPIC-pn count rate in the soft band (0.2--1.0\,keV, corrected for vignetting) was determined by the source detection algorithm to (0.21$\pm$0.01) \cts.
We extracted the events to produce images, the spectrum, and the light curve using the SAS task \texttt{evselect}. For the extraction of the source products, circular regions around the source 
position and a nearby source-free area were used as source and background regions (with radii of 40\arcsec\ and 60\arcsec, respectively, see Fig.~\ref{fig:mosaic}). Due to the higher sensitivity of the pn detector compared to the MOS detectors 
at low energies, we used only EPIC-pn data for spectral analysis. Single- and double-pixel events (PATTERN 0--4) were selected,
excluding known bad CCD pixels and columns (FLAG 0).
For the spectra, we removed times of increased flaring activity when the background was above a threshold of 8 counts ks$^{-1}$ arcmin$^{-2}$ (7.0--15.0\,keV band).
 
The SAS tasks \texttt{arfgen} and  \texttt{rmfgen} were used to generate the corresponding detector response files for the EPIC-pn spectrum. 

We obtained a net exposure of 25.7\,ks for EPIC-pn and EPIC-MOS, respectively, after we removed intervals of high background flaring activity, which occurred only at the end of the observation. Table\,\ref{tabobsero} summarises the observation details.
\begin{table*} 
\centering
\caption{\ero and \xmm observations of \esrc} 
\label{tabobsero} 
\begin{tabular}{cccc} 
\hline\hline\noalign{\smallskip}
Observation\tablefootmark{a} & Obs. time & Net exposure\tablefootmark{b} & count rate\tablefootmark{c} \\
          &   T$_{start}$ -- T$_{stop}$ (UTC) & ks & \cts  \\
\noalign{\smallskip}\hline\noalign{\smallskip}
eRASS1 & 2020-03-13 12:57:12 -- 2020-03-28 00:57:28 & 1.3 &  0.34$\pm0.02$  \\ 
 eRASS2 & 2020-09-15 22:02:40 -- 2020-10-02 18:02:40 & 1.5 &  0.34$\pm0.02$ \\
 eRASS3 & 2021-03-10 13:57:08 -- 2021-03-24 17:57:25 & 1.1 &  0.36$\pm0.02$  \\
 eRASS4 & 2021-09-11 14:02:46 -- 2021-09-27 22:02:39 & 1.3 &  0.29$\pm0.01$ \\
 \xmm & 2022-01-21 16:06:42 -- 2022-01-22 01:44:24 & 25.7 &  0.21$\pm0.01$  \\
\noalign{\smallskip}\hline
\end{tabular} 
\tablefoot{
\tablefoottext{a}{eRASSn denotes the \ero\, survey number}
\tablefoottext{b}{Net exposure after correcting for vignetting and normalised to seven telescope modules. } \tablefoottext{c}{Count rate in the 0.2--8\,keV energy band.}}

\end{table*}

\section{X-ray data analysis}
\label{sec:xanalysis}

\subsection{Temporal analysis}
\label{sec:temporal}

\begin{figure*}[ht]
\centering
   \resizebox{0.48\hsize}{!}{\includegraphics{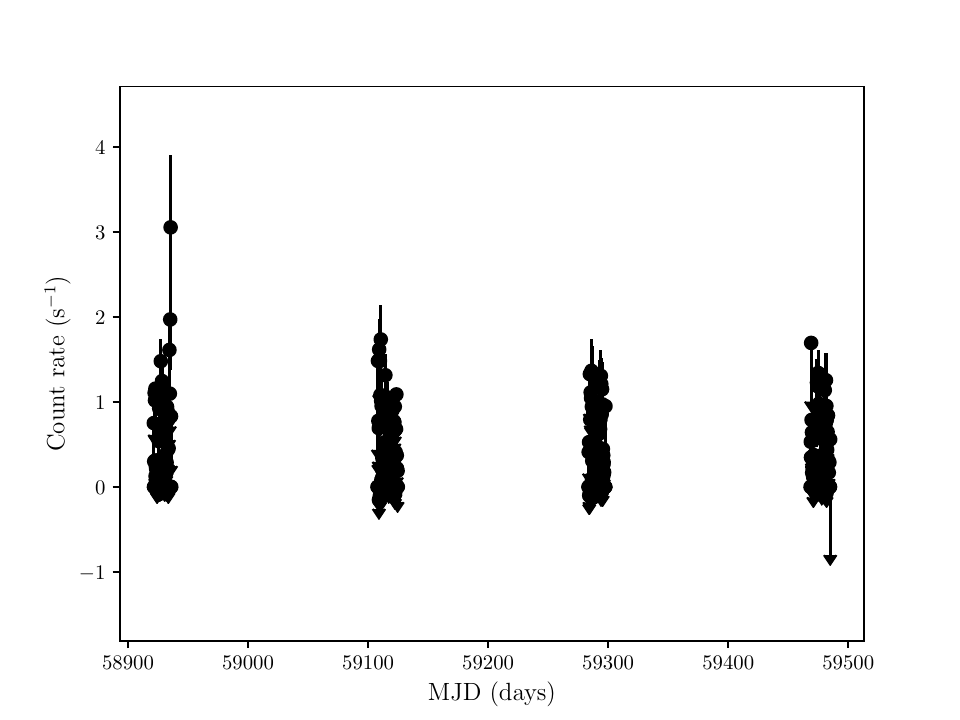}}
   \resizebox{0.48\hsize}{!}{\includegraphics{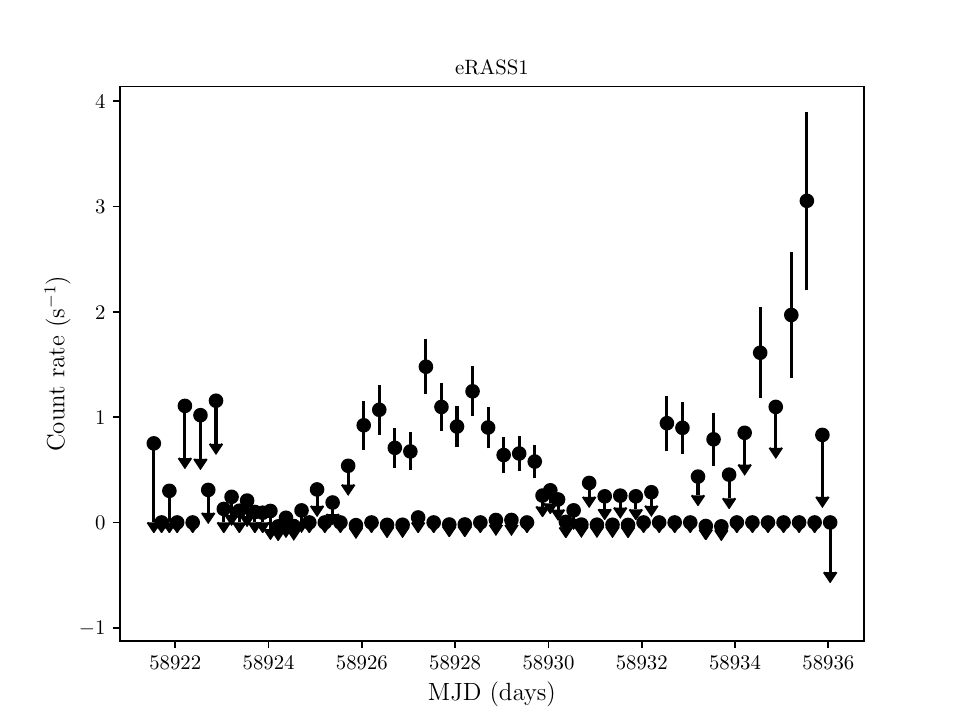}} 
\caption{\ero 0.2--2\,keV light curves of \esrc. The black points represent a single scan. The count rates of bins with fewer than ten counts are plotted as 1\,$\sigma$ upper limits.  The total light curve of eRASS1--4 is shown left, with a zoomed-in view covering eRASS1 on the right.}
  \label{fig:eRO_LC}
\end{figure*}

\begin{figure*}
\centering
  \resizebox{0.48\hsize}{!}{\includegraphics{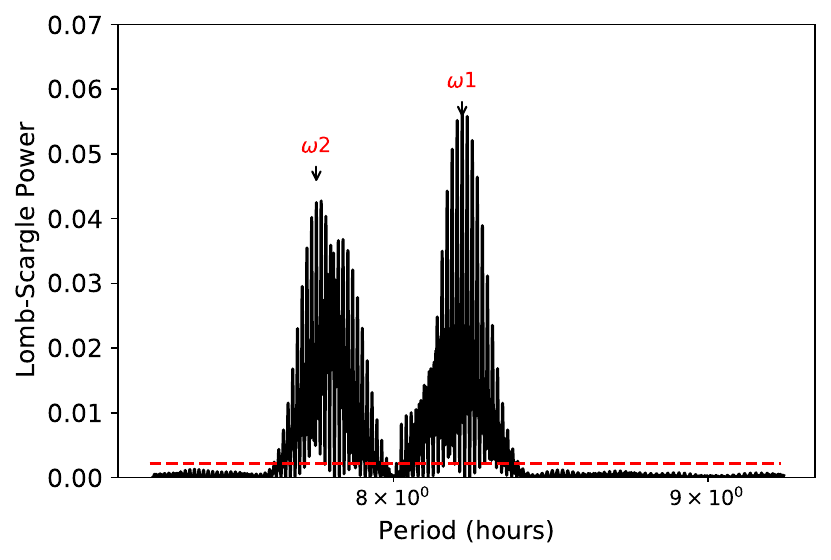}}
  \resizebox{0.48\hsize}{!}{\includegraphics{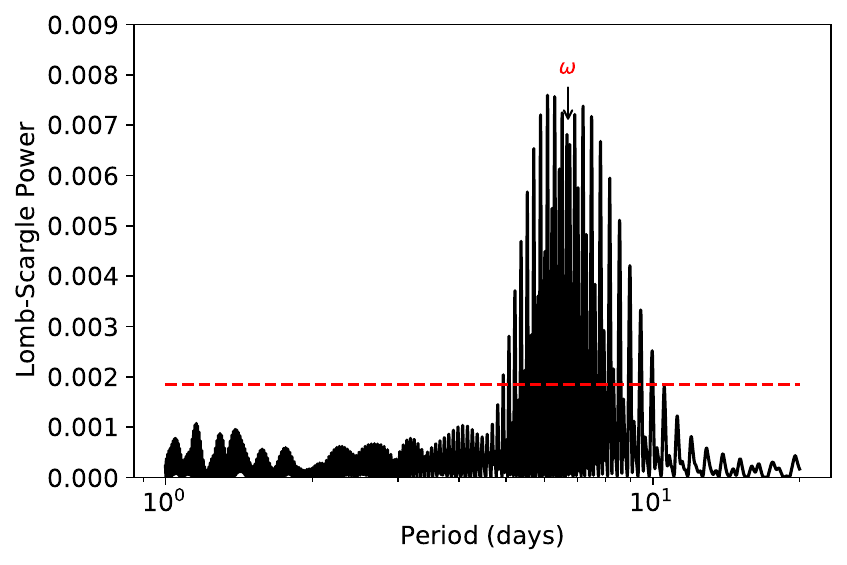}}
\caption{Lomb-Scargle periodograms of the eRASS1--4 light curve showing the periodicities at 7.77 and 8.21 hours (left) and at 6.7 days (right). The dashed red line marks the 99\% confidence level.}
  \label{fig:eRO_LC_LS}
\end{figure*}

The eRASS surveys scanned the whole sky in great circles with a scanning period of 4 hours, the length of an eRODay that intercepted at the ecliptic poles. Although a typical scan lasts for up to 40\,s (separated by 4\,hours), when a source is located in the direction of the LMC, which is close to the South Ecliptic Pole, it is scanned many times during a visibility period of several weeks per eRASS. 
Each \ero survey (eRASSn) is repeated after six months, allowing a source to be monitored on timescales of years. We present here the \ero data of \esrc from the first four complete all-sky surveys (eRASS1, 2, 3, and 4).
Figure~\ref{fig:eRO_LC} shows the 0.2--2\,keV \ero light curve of \esrc as it was scanned during each eRASS.
The source exhibited a stable count rate (see Table\,\ref{tabobsero}, after background subtraction, taking into account corrections for vignetting and losses due to the point spread function of the telescopes) in all four surveys. The light curves show a dipping behaviour alternating between “on” and “off” (zero-flux) states in each consecutive scan. An additional longer-term modulation is also apparent (see Fig.~\ref{fig:eRO_LC} right). In order to investigate the possible periodic signals in detail, we used a Lomb-Scargle (LS) periodogram analysis \citep{1976Ap&SS..39..447L,1982ApJ...263..835S}. Periodicities at two different timescales are clearly detected (Fig.~\ref{fig:eRO_LC_LS}); the alternating dipping feature corresponds to a main peak at 8.21\,hours and a secondary one at 7.77\,hours. A second periodicity is also detected at 6.7\,days, corresponding to the longer-term modulation in the light curve. 

\begin{figure}
\centering
  \resizebox{\hsize}{!}{\includegraphics{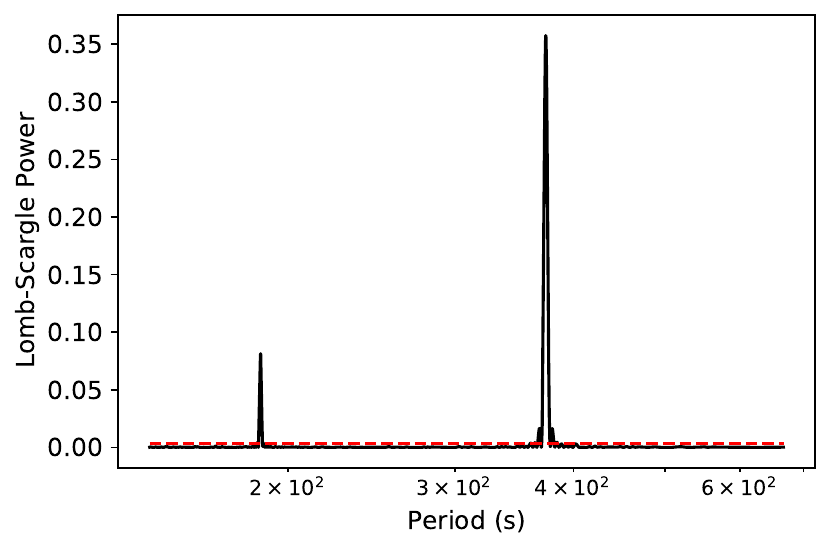}}
\caption{
    Lomb-Scargle periodogram of \esrc obtained from the combined \xmm data (0.2--1.0\,keV). Pulsations are detected with of period of 374\,s. The dashed red line marks the 99.73\% confidence level obtained by the block-bootstrapping method.}
  \label{fig:PNpower}
\end{figure}

\begin{figure}
  \centering
  \resizebox{0.9\hsize}{!}{\includegraphics[]{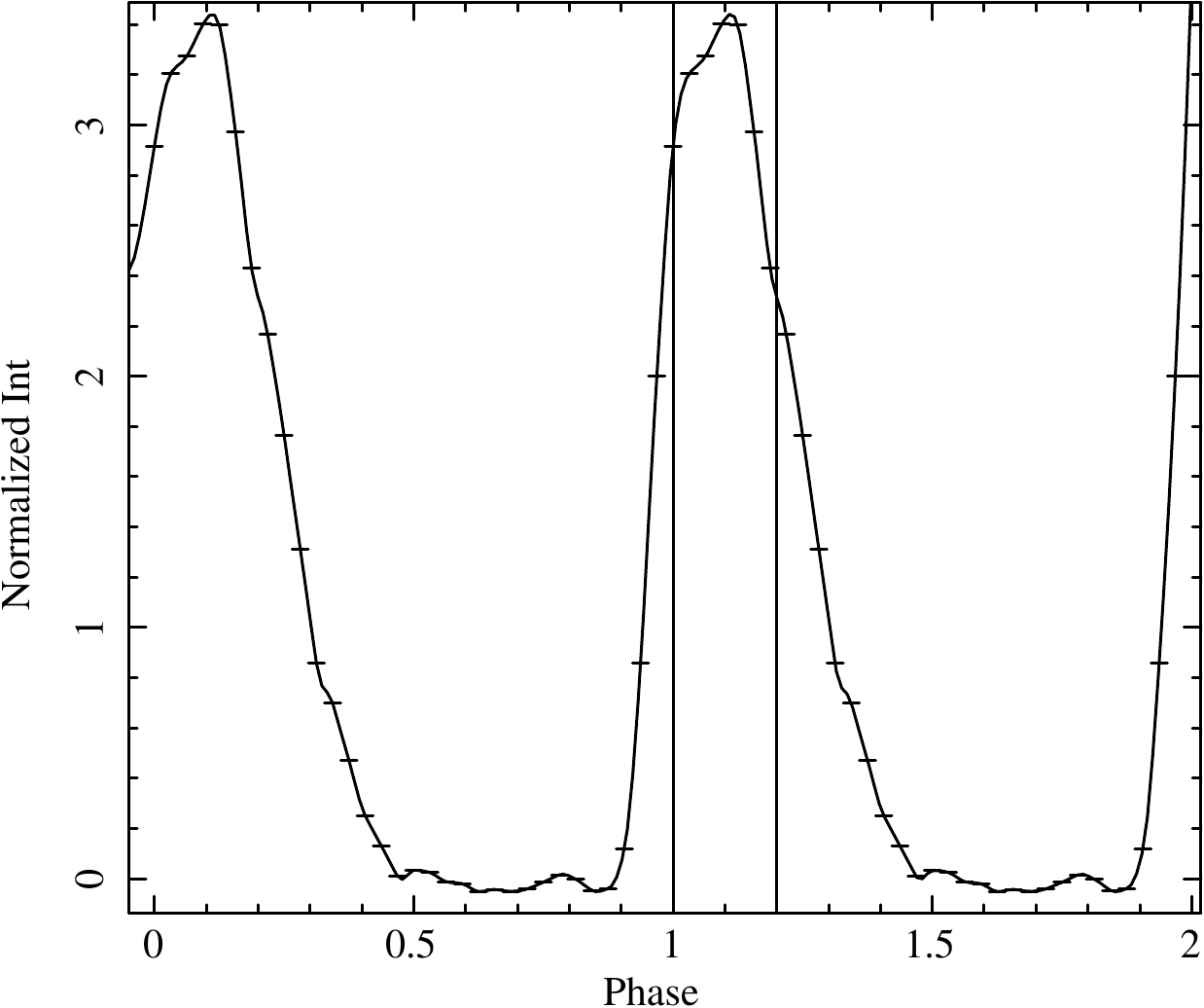}}
  \caption{
    Background-subtracted and combined \xmm light curve folded with the best-fitted period showing the pulse profile of \esrc in the energy band of 0.2--1\,keV. The pulse profile is normalised to the mean count rate of  0.21\,\cts. The vertical lines show the interval taken as on-pulse for phase-resolved spectroscopy.
  }
  \label{fig:pp}
\end{figure}

The background-subtracted 0.2--2.0\,keV light curves of \esrc obtained from the three EPIC instruments do not exhibit any long-term variability on timescales of hours. However, a Lomb-Scargle periodogram reveals a very strong signal at 374\,s as shown in Fig.~\ref{fig:PNpower}.
 
The best-determined period and the 1\,$\sigma$ uncertainty are given by 374.03$\pm$0.21\,s. 

To estimate the uncertainty on the periodic signal, we applied a block-bootstrapping method, similar to that described in \citet{1999ApJ...522L..49G} and \citet{2022A&A...661A..20C}, by generating a set of 10\,000 light curves. The background-subtracted and combined \xmm light curve obtained by combining the two modules and folding with the best-obtained period is shown in Fig.~\ref{fig:pp}.
The X-ray signal is characterised by a 100\% modulation with a duty cycle of $\sim$50\%. 

The periodicity detected with \xmm further indicates that the 8.2\,hours and 6.67\,days are aliases due to sampling of the \ero light curve with the \ero scanning period of 4\,hours, which is almost exactly 38.5 times the pulse period of 374.03 s.
Therefore, \ero detects alternating on-off states every other scan because of the pulse profile, which is $\sim$50\% on-off. Also, 6.67 days corresponds very accurately to 40 \ero scans. 

\subsection{Period constraints from \ero survey data}
\label{sec:temporalEROS}

\begin{figure}
  \centering
  \resizebox{0.9\hsize}{!}{\includegraphics[]{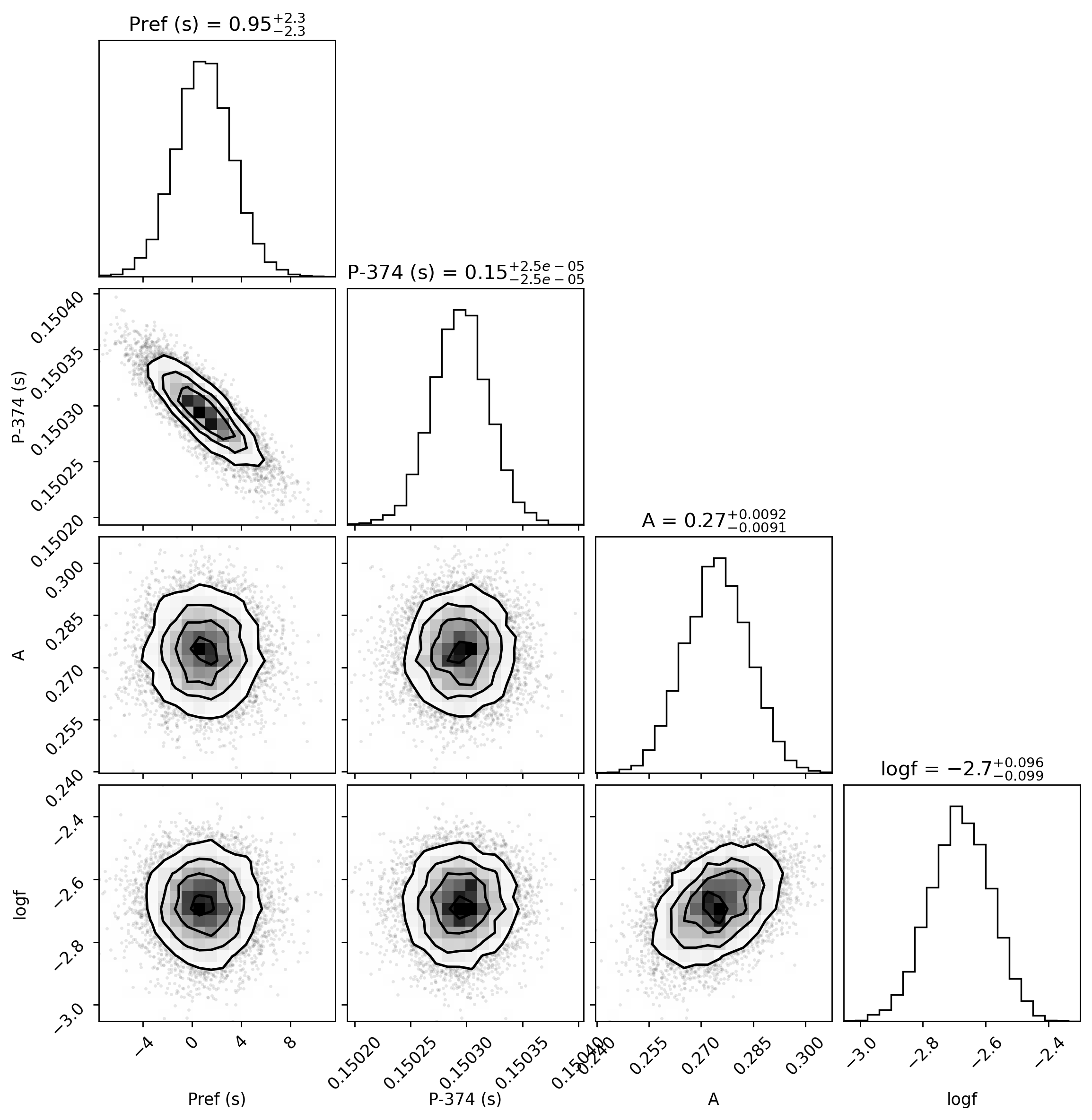}}
  \caption{Corner plot of the model fitted to the \ero data. $P_{\rm ref}$ is the reference time compared to the time of the first \ero point of the data set, i.e. MJD 58921.70648. 
  }
  \label{fig:alias_ER_corner}
\end{figure}

\begin{figure}
  \centering
  \resizebox{0.9\hsize}{!}{\includegraphics[]{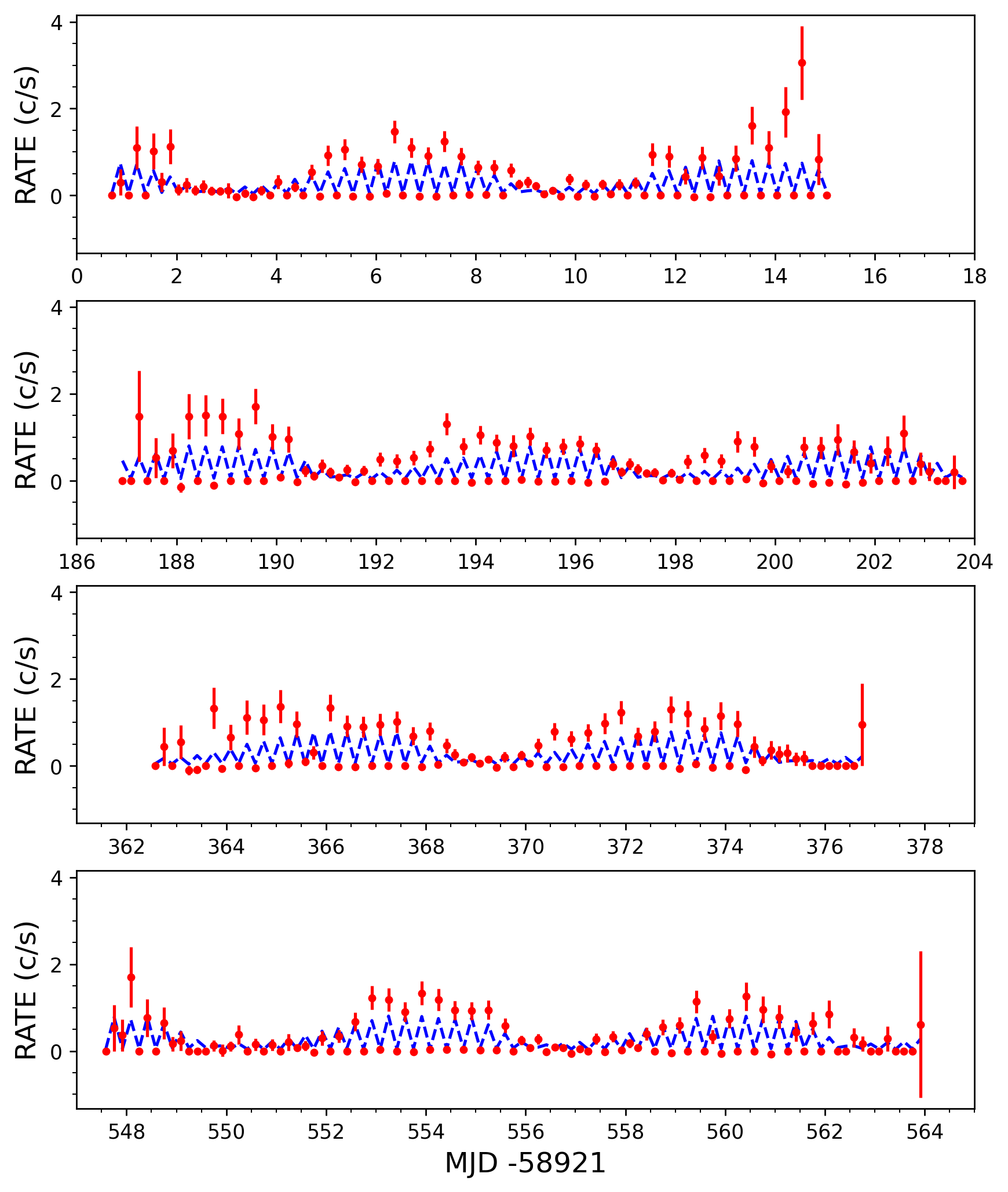}}
  \caption{Comparison of the \ero data (red points) with the predictions of the aliasing model (dashed blue line). For plotting purposes we have connected the model points, although this bears no physical meaning.
  }
  \label{fig:alias_ER}
\end{figure}

As mentioned in the previous section, the on-off pattern observed in the \ero light curve is a result of the aliasing between the actual pulse period and the $\sim$4-hour sampling of \ero. In order to investigate this, we constructed a so-called aliasing model based on the “pulse” profile observed by \xmm and the \ero sampling and used it to reconstruct a fake \ero light curve. For simplicity we assumed each model point is based on a single point of the \xmm profile, that is, we did not integrate for \ero exposure. This model can reconstruct the 6.7\,d periodicity. However, the resulting period is quite sensitive to the actual pulse period, as a decrease of 0.1\,s in the period can increase the aliasing period by a factor of two. This dependence can be used to constrain the periodicity of \esrc by fitting the alias model to the actual \ero light curve. 
We used a python implementation of the Goodman \& Weare’s Affine Invariant Markov chain Monte Carlo (MCMC) Ensemble sampler (i.e. {\tt emcee}). 
Our model has four free parameters, the reference phase, period, and amplitude (i.e. re-normalisation between \xmm and \ero) of the periodic signal, while we also added a term of $ln{f}$ to account for the systematic scatter and noise of our data not included in the statistical uncertainties of the measurements. 
We derived a period of 374.15029(3) s with the corner plot of the posterior distribution of the model parameters shown in Fig.~\ref{fig:alias_ER_corner}.
The comparison of the model and the data is shown in Fig.~\ref{fig:alias_ER}.

To further investigate the uncertainties of this method we performed additional tests. We first repeated the fit in the individual \ero epochs and found a consistent period with an uncertainty of the order of 0.003 s.
By repeating the procedure with and without barycentric corrections on the complete \ero data set we found a difference of 5$\times10^{-5}$ s and comparable uncertainties. Another potential question is the presence of a period derivative. Including a period derivative in the model, we found a slightly different period of 374.1506(2)\,s and a derivative of 1.3(5)$\times10^{-11}$ s s$^{-1}$. The uncertainties correspond to 1$\sigma$ intervals. 
We note that, although the error is quite small, the real uncertainty should be dominated by uncertainties  related to the barycentric corrections (in the absence of an absolute timing reference for \ero and spacecraft clock drifts in the current CALDB version), and multi-modal solutions in the parameter space. 
When exploring the parameter (i.e. $P$,\,$\dot{P}$) space more thoroughly, we found degenerate solutions with both positive and negative spin period derivatives. Since the problem has a multi-modal parameter space we also attempted an alternative strategy of phase-connecting the \ero and \xmm data sets. A reasonably good phase connection can be obtained with a constant $P$ of 374.1510(2) s, with an upper limit on $\dot{P}$ $\leq3.9\times10^{-12}$ s s$^{-1}$\,.
While a solution with a positive period derivative as quoted above (i.e. 1.3(5)$\times10^{-11}$ s s$^{-1}$\,) yields a similar result, obtaining a solution with a negative derivative affects the phase connection of the aliasing model in the \ero data, and thus seems less likely.
Based on the above experiments, we can conclude a few things. First, the presence of a stable aliasing pattern in the \ero data excludes a significant period evolution. Moreover, the multiple tests in the \ero data have revealed that we can measure the period of \esrc more accurately than the original \xmm measurement, and we can conclude that the period can be determined to 374.1503(5) s, considering all uncertainties.

\subsection{Spectral analysis}
\label{sec:spectral}

In order to model the spectra from the EPIC-pn observation and the eRASS, we used a Bayesian methodology \citep[see][\texttt{BXA}]{2014A&A...564A.125B}\footnote{\url{https://github.com/JohannesBuchner/BXA}}, which allowed us to robustly model the low signal-to-noise eRASS spectrum as well as to look for additional spectral components in the case of EPIC-pn. \texttt{BXA} connects the nested sampling \citep{2004AIPC..735..395S} algorithm MultiNest \citep{2009MNRAS.398.1601F} with \xspec. It explores the parameter space and can be used for parameter estimation (probability distributions of each model parameter and their degeneracies) and for model comparisons (computation of Bayesian evidence, $Z$).

We modelled the background spectrum (extracted from the corresponding background regions) and used the best-fit models with an area scaling factor to account for the background component in the spectrum from the source region. We use the automatic background-fitting method described in the appendix of \citet{2018A&A...618A..66S}, which is implemented in \texttt{BXA}. In this method, the background spectrum is modelled phenomenologically as a function of detector channels after a log(1 + counts) transformation. Principal component analysis (PCA) is run on the unbinned background spectra. The first six principal components (PCs) are then linearly combined to fit the particular background spectrum of interest.
Starting from the mean spectrum, PCs are iteratively added as long as the Akaike information criterion \citep[AIC;][]{1974ITAC...19..716A} of the fit is significantly improved.
After finding the linear combination of PCs that describes the spectrum best, Gaussian lines are added to further model detector-related features. These added Gaussians can model features that might appear in some individual spectra and were missed by the PCA. 
The \xmm and \ero\,background models could be sufficiently described by five PCs and two PCs, respectively. The addition of Gaussian components did not further improve the AIC. Finally, the best-fit background model spectrum was converted into an \xspec table model with a scale parameter (the ratio between the BACKSCAL of the source and the background when fitting the source spectrum). A similar analysis was conducted in \citet{2022A&A...661A...5L}.

The \ero spectra can be fitted with an absorbed black-body model. To account for the \nh,
first, we assumed that the source is located in the LMC and used two column densities along the line of sight.
One accounts for the Galactic foreground with solar abundances according to \citet{2000ApJ...542..914W} and was fixed at the value obtained from \Hone measurements \citep{1990ARA&A..28..215D}\footnote{Extracted using NASA's HEASARC web interface \url{https://heasarc.gsfc.nasa.gov/cgi-bin/Tools/w3nh/w3nh.pl}}. The other (free in the fit), with metal abundances set to 0.5, reflects the absorption by the interstellar medium of the LMC \citep{2002A&A...396...53R} and local to the source. Second, if the source could be closer and located in the Milky Way, we used only one column density with the solar abundance, allowing it to be a free parameter in the spectral fit. Luminosities were corrected for absorption and calculated assuming a distance of 50\,kpc in the case of the LMC and 1\,kpc in the case of our Galaxy.

At first, \ero spectra from the different epochs (eRASS1/2/3/4) were modelled separately to investigate possible variation in spectral parameters during the different surveys. We however verified that the spectral parameters were consistent with each other and present the spectrum combined from all four eRASSs (eRASS:4) with data summed from the TMs with an on-chip filter (TM1, 2, 3, 4, and 6). 
 
\begin{figure}
  \centering
  \resizebox{0.9\hsize}{!}{\includegraphics[]{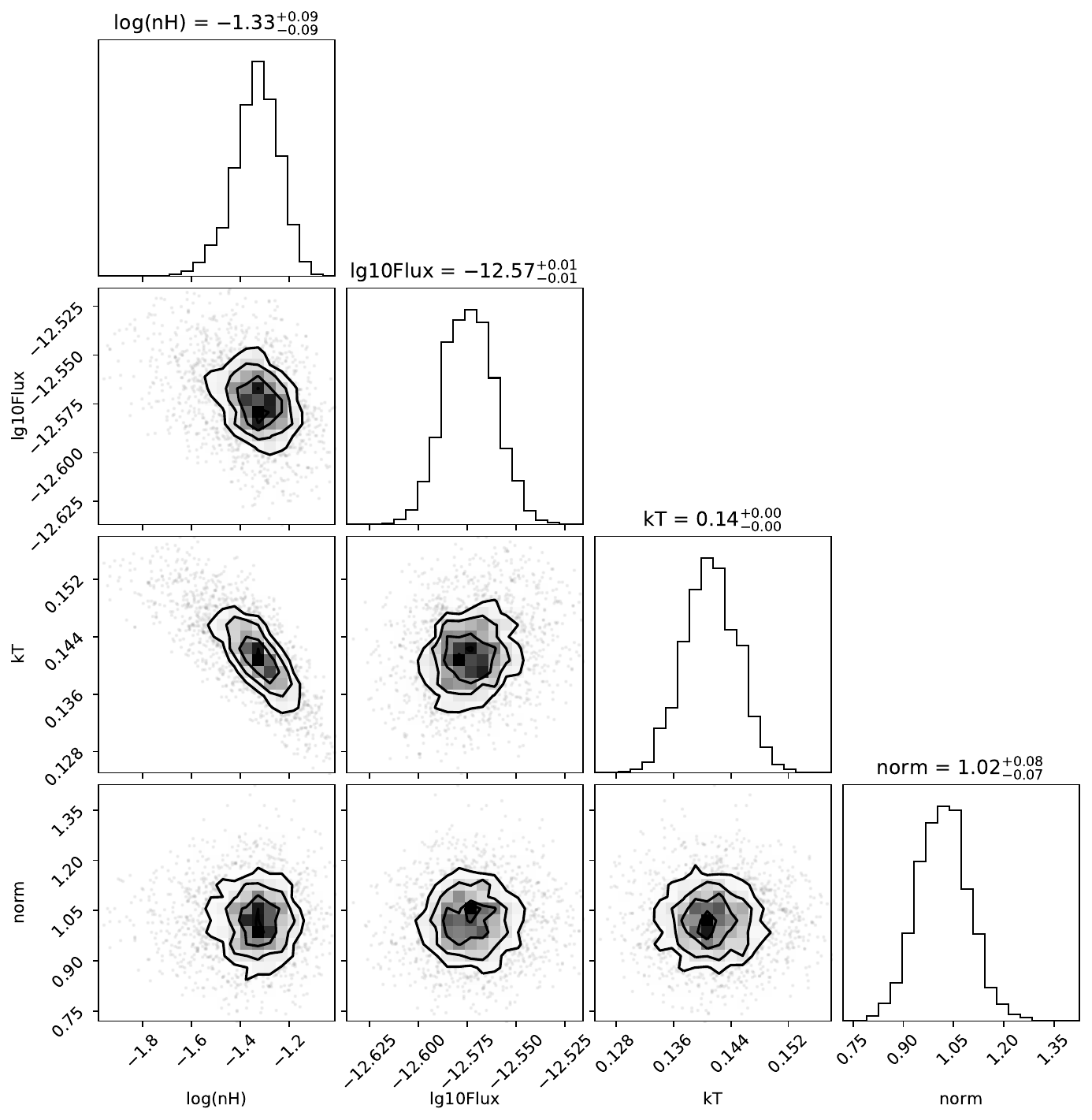}}
  \caption{Corner plot of the posterior distributions for the eRASS:4 spectral fit with a single \nh model. \nh is in units of $10^{22}$ atoms cm$^{-2}$}.
  
  \label{fig:corner-ero}
\end{figure}

 From the posterior distribution of each parameter, we measured the median and the 1\,$\sigma$ percentile confidence interval around the median as shown in the corner plots in Fig.~\ref{fig:corner-ero}. The posterior distributions of the model parameters overlayed with the convolved spectrum are shown in Fig.~\ref{fig:eRO_spec}.

The spectral parameters are detailed in Table\,\ref{tab:spectral}, where the $\log Z$ column (normalised to highest) shows the computed evidence.

In the case of the \xmm/EPIC spectrum, we used the same model of a black-body emission attenuated by photo-electric absorption as in the case of \ero. However, we additionally tested for the presence of a bremsstrahlung model \citep[bremsstrahlung is generally seen in the X-ray spectra of mCVs, see e.g.][]{2017PASP..129f2001M}.
We fixed the temperature at 10\,keV and assumed that black-body and bremsstrahlung components are attenuated by the same \nh. For this, we computed the difference in the $\log Z$ that corresponds to a Bayes factor (BF), which can be used to discriminate between the models. A commonly used way to interpret the BF values is the Jeffrey scale, which strengthens the choice of one model over the other approximately every time that the logarithm of the BF increases by one in natural logarithmic units \citep[]{10.1214/09-STS284}; see also \citet[][]{2014A&A...564A.125B}. Uniform model priors are assumed in this case, with a cut of $\log _{10}$(30) considered to be “decisive”.
 
\begin{figure}
  \centering
  \resizebox{0.9\hsize}{!}{\includegraphics[]{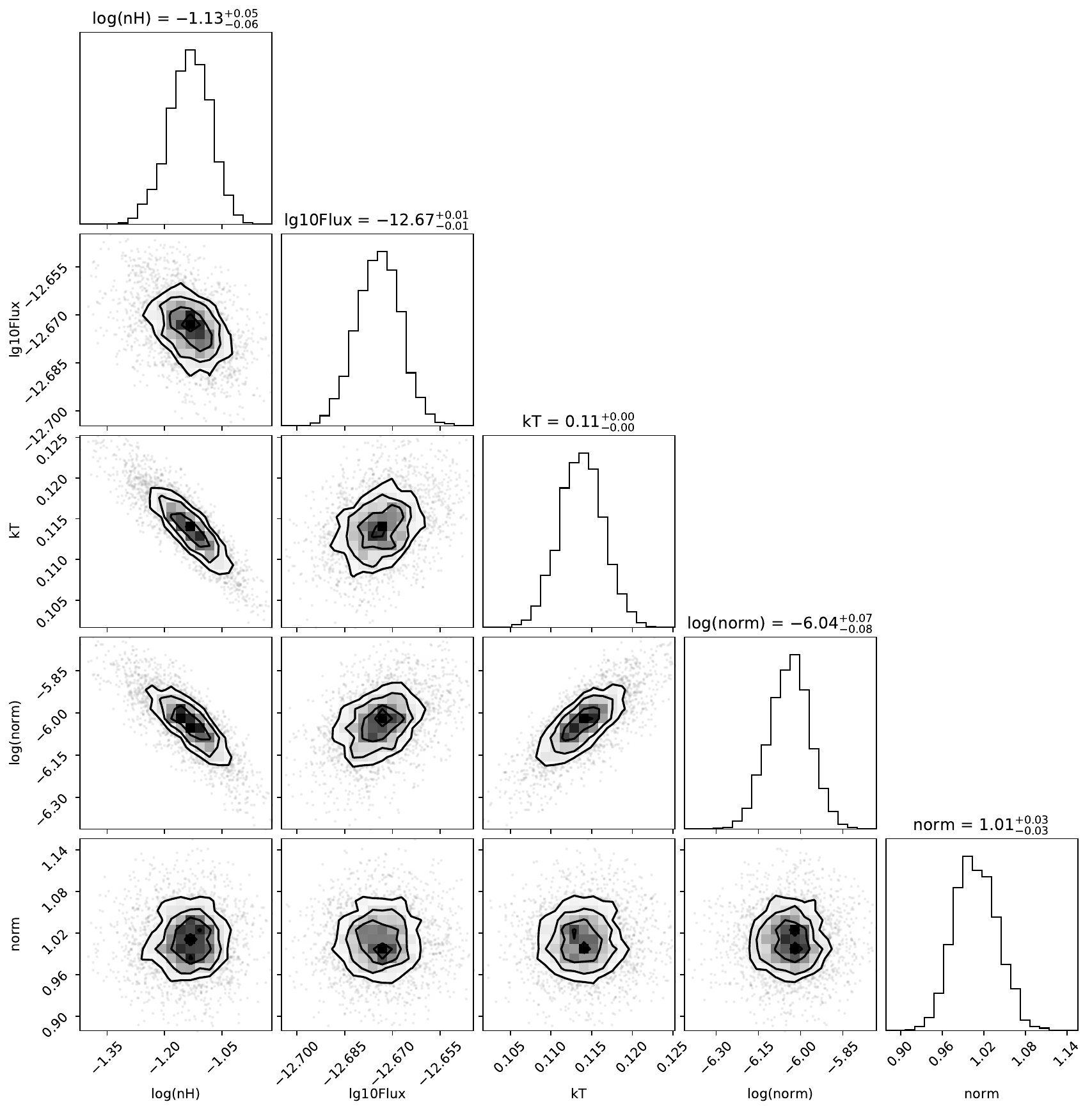}}
  \caption{Corner plot of the posterior distributions for the \xmm EPIC-pn spectral fit with a single \nh and after adding the bremsstrahlung component. \nh is in units of $10^{22}$ atoms cm$^{-2}$}.
  
  \label{fig:corner-xmm}
\end{figure}

\begin{figure}

\resizebox{\hsize}{!}{\includegraphics{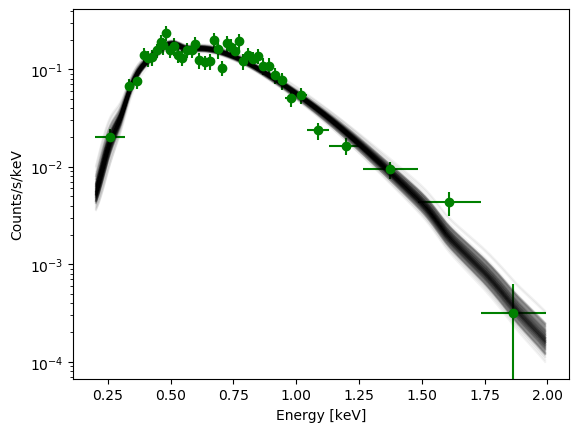}}
\resizebox{\hsize}{!}{\includegraphics{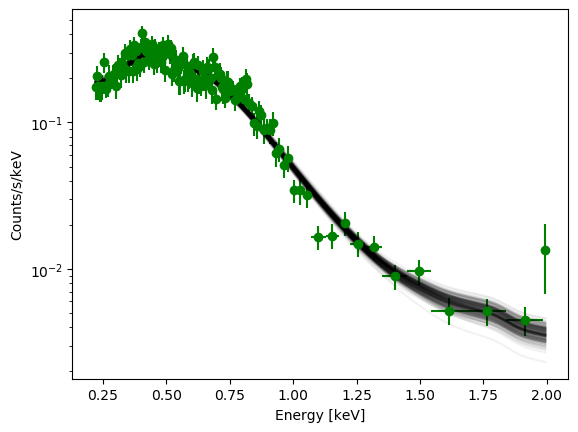}}
  \caption{Spectral fit of \esrc using the \ero spectrum from eRASS:4 (top) and EPIC-pn (bottom) together with the posterior distributions of model parameters. The spectrum for each eRASS was extracted by combining data from the TMs with an on-chip filter. See Table\,\ref{tab:spectral} for the model parameters. Darker and lighter bands enclose the 68 \% and 95 \% posterior uncertainties on the model at each energy.}

  \label{fig:eRO_spec}
\end{figure}

The addition of the bremsstrahlung component showed a significant improvement in this regard with a $\log Z$ of $-$62.7. The contribution of the bremsstrahlung component corresponds to an \lx$\simeq$4\ergs{32}
in the 2--8\,keV range.
The  median and the 1\,$\sigma$ percentile confidence intervals for each parameter are shown in the corner-plots in Fig.~\ref{fig:corner-xmm}. The posterior distributions of the model parameters overplotted with the convolved spectrum are shown in Fig.~\ref{fig:eRO_spec}, and the spectral parameters are summarised in Table\,\ref{tab:spectral}.

We also extracted the spectra at the “on” and “off” pulse phases, as shown in Fig.~\ref{fig:pp}. The bremsstrahlung component is detected more
significantly in the “off” pulse spectrum, as seen in Table\,\ref{tab:spectral}. The rest of the parameters are consistent within their errors.

\begin{table*}
\centering
\caption[]{Spectral fit results of \esrc from BXA.
The first set of the table corresponds to the fit parameter “in the LMC”
interpretation (two column densities) and the second set (lower in the table)
corresponds to the “foreground” interpretation.}
\begin{tabular}{lllccccccccc}
\hline\hline\noalign{\smallskip}
\multicolumn{1}{l}{Obs} &
\multicolumn{1}{c}{kT$_{\rm bb}$} &
\multicolumn{1}{c}{Norm} &
\multicolumn{1}{c}{N$_{\rm H}^{\rm Gal}$} &
\multicolumn{1}{c}{N$_{\rm H}^{\rm LMC}$} &
\multicolumn{1}{c}{$\log Z$} &
\multicolumn{1}{c}{dof} &
\multicolumn{1}{c}{cstat/dof} &
\multicolumn{1}{c}{F$_{\rm observed}$\tablefootmark{a}} &
\multicolumn{1}{c}{L\tablefootmark{b}} &
\multicolumn{1}{c}{R$_{\rm BB}$} \\
\multicolumn{1}{l}{--} &
\multicolumn{1}{c}{(eV)} &
\multicolumn{1}{c}{brems} &
\multicolumn{1}{c}{(\ohcm{21})} &
\multicolumn{1}{c}{(\ohcm{21})} &
\multicolumn{1}{c}{--} &
\multicolumn{1}{c}{--} &
\multicolumn{1}{c}{--} &
\multicolumn{1}{c}{(erg cm$^{-2}$ s$^{-1}$)}&
\multicolumn{1}{c}{(erg s$^{-1}$)} &
\multicolumn{1}{c}{(km)} \\
\noalign{\smallskip}\hline\noalign{\smallskip}
  eRO    &  $131^{+3}_{-2}$ & --              & 0.88 & $<0.5$ & $-$293.3 & 249 & 1.5 & $2.5^{+0.1}_{-0.2}$    &  $1.4 \times 10^{35}$ & 70.1$^{+3.6}_{-3.4}$ \\
  \noalign{\smallskip}
  EPIC   &  $109.5\pm1.4$ & $1.9\pm1.0$ & 0.88 & $<0.6$ & $-$418.2 & 353 & 1.2 & $2.09^{+0.03}_{-0.05}$ &  $1.7 \times 10^{35}$ & 96.8$^{+3.6}_{-3.5}$ \\
  \noalign{\smallskip}
   on    & $109.4\pm1.5$ & $1.4\pm1.0$  & 0.88 & $<0.6$ & $-$366.9 & 353 & 0.9 & $5.25^{+0.15}_{-0.06}$ &  $5.6 \times 10^{35}$ & 153.4$^{+3.3}_{-4.8}$ \\
  \noalign{\smallskip}
  off &   $110.2\pm1.5$ & $2.8\pm1.0$   & 0.88 & $<0.8$ & $-$417.6 & 353 & 1.1 & $1.47^{+0.21}_{-0.42}$ &  $1.4 \times 10^{35}$ & 75.2 $^{+4.4}_{-3.9}$ \\
 
\noalign{\smallskip}\hline\noalign{\smallskip}
  eRO &  $141^{+4}_{-5}$ & --          & 0.47$^{+0.01}_{-0.01}$  & --    & $-$283.5 & 249 & 1.3 & $2.6^{+0.1}_{-0.2}$ &  $5.9 \times 10^{31}$ &  1.03 $^{+0.09}_{-0.08}$ \\
  \noalign{\smallskip}
  EPIC  & $113.8\pm2.7$  & $2.1\pm1.0$ & 0.74$^{+0.02}_{-0.01}$& -- &$-$418.7 & 353 & 1.1 &  $2.11^{+0.04}_{-0.03}$&  $7.3 \times 10^{31}$ &  2.0 $^{+0.2}_{-0.3}$ \\
  \noalign{\smallskip}
  on  & $112.8\pm3.3$    & $1.5\pm1.0$ & 0.77$^{+0.02}_{-0.03}$  & --    &  $-$368.8 & 353 & 0.9 & $5.31^{+0.11}_{-0.12}$ &  $2.1 \times 10^{32}$ & 2.0$^{+0.4}_{-0.2}$ \\
  \noalign{\smallskip}
  off &  $116.1\pm4.1$   & $3.4\pm1.0$  & 0.68$^{+0.02}_{-0.02}$ &  -- & $-$418.5 & 353 & 1.0 & $1.41^{+0.4}_{-0.3}$ & $5.8 \times 10^{31}$ & 0.5$\pm0.3$ \\
\noalign{\smallskip}\hline

\end{tabular}
\tablefoot{
Best-fit parameters using a model with absorbed black-body and bremsstrahlung emission (in the case of \xmm). The normalisation of the bremsstrahlung component is in units of $10^{-3}$. The bremsstrahlung component is not well constrained, and the temperature was fixed at 10\,keV (see Sect.\,\ref{sec:spectral}). The $\log Z$ indicates the goodness of the fit.
Errors indicate 1\,$\sigma$ percentile values. Corresponding cstat/dof values are also quoted to get an independent estimate of the fit quality. Errors indicate 1\,$\sigma$ percentile values. 

\tablefoottext{a}{Fluxes are provided for the 0.2--2.0\,keV band in units of $10^{-13}$ to allow a comparison with values published for SSSs based on \rosat observations.}
\tablefoottext{b}{Source luminosities (0.2--2.0\,keV) corrected for absorption, assuming a distance of 50\,kpc \citep{2013Natur.495...76P} in the upper part of the table. The Galactic foreground column density was taken from \citet{1990ARA&A..28..215D}.
For the model in the lower part of the table, only a Galactic absorption component (with free column density in the fit) and a distance of 1\,kpc was assumed.
A note of caution regarding absorption-corrected X-ray luminosities in soft energy bands: column densities with large errors lead to a large uncertainty on the luminosity. Norm denotes the normalisation of the bremsstrahlung component.}
}
\label{tab:spectral}
\end{table*}

\section{UV, optical, and near-infrared data}
\label{sec:optical}

Using the VizieR catalogue access tool{\footnote{\url{http://cdsarc.u-strasbg.fr/viz-bin/VizieR}}}, a star with NUV$\sim$22 mag was found marginally consistent within the 1\,$\sigma$ error circle of the \xmm position of \esrc in the GALEX DR5 \cite[][]{2011Ap&SS.335..161B} catalogue (see Table\,\ref{tab:grond_mag}). 
 
 To obtain deeper images of the area around \esrc, we performed observations with the Gamma-ray Burst Optical Near-IR Detector 
\citep[GROND,][]{2008PASP..120..405G} at the MPG 2.2\,m telescope in La Silla, Chile on August 4 2016. 
For g', r', i', and z' we obtained a total of 21.2\,ks in each filter, while for J, H, and K 19.4\,ks per filter were accumulated.
We analysed the data with the standard tools and methods described in \citet{2008ApJ...685..376K}.
Photometric calibration for the g', r', i', and z' filter bands was obtained from the observation of an SDSS (Sloan Digital Sky Survey) standard star field. 
The J, H, and K photometry was calibrated using selected 2MASS stars \citep{2006AJ....131.1163S}. 
However, we note that the GROND photometry was contaminated by a nearby bright object, and therefore we did not use it to analyse the spectral energy distribution (SED) of the object.

The optical counterpart was also identified in archival data from
\decam \citep[][]{2015AJ....150..150F}, a wide-field CCD camera on the Blanco 4-m telescope at the Cerro Tololo Inter-American Observatory. 24 exposures in g, r, i, and z filters from three programmes were used (including the \decam \ero Survey, DeROSITAS, an optical companion survey to the \ero survey).
The exposures were instrumentally calibrated by the \decam Community Pipeline \citep[][]{DCP} and catalogued for the Legacy Surveys' \citep{2019AJ....157..168D} data release DR10.
The analysis uses the {\texttt Tractor} \citep{2016ascl.soft04008L} forward modelling method. The method fits models to the observations, taking the point spread functions from the exposures into account. The {\texttt Tractor} fitting agrees with a very compact galaxy or a point source; that is to say, the data are consistent with either a slightly extended source or a star.  The position of the optical counterpart corresponds to $\alpha_\mathrm{J2000.0}= 06\rahour\,08\ramin\,38\fs88$ and $\delta_\mathrm{J2000.0} = -70\degr\,40\arcmin\,14\farcs52$
with a 1\,$\sigma$ statistical uncertainty of $0\farcs003$ and a remaining systematic error of a similar order after astrometric correction\footnote{https://www.legacysurvey.org/dr8/external/}.

\begin{figure*}
\centering
\resizebox{0.6\hsize}{!}{\includegraphics{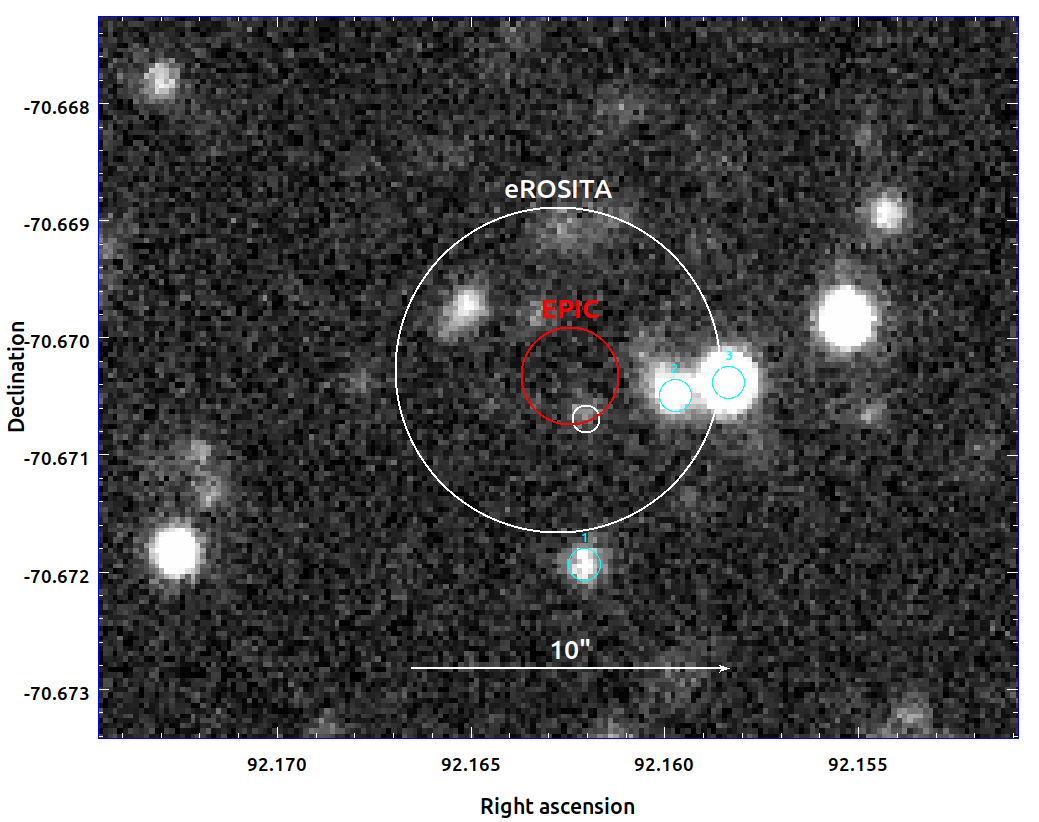}}
 \resizebox{0.52\hsize}{!}{\includegraphics{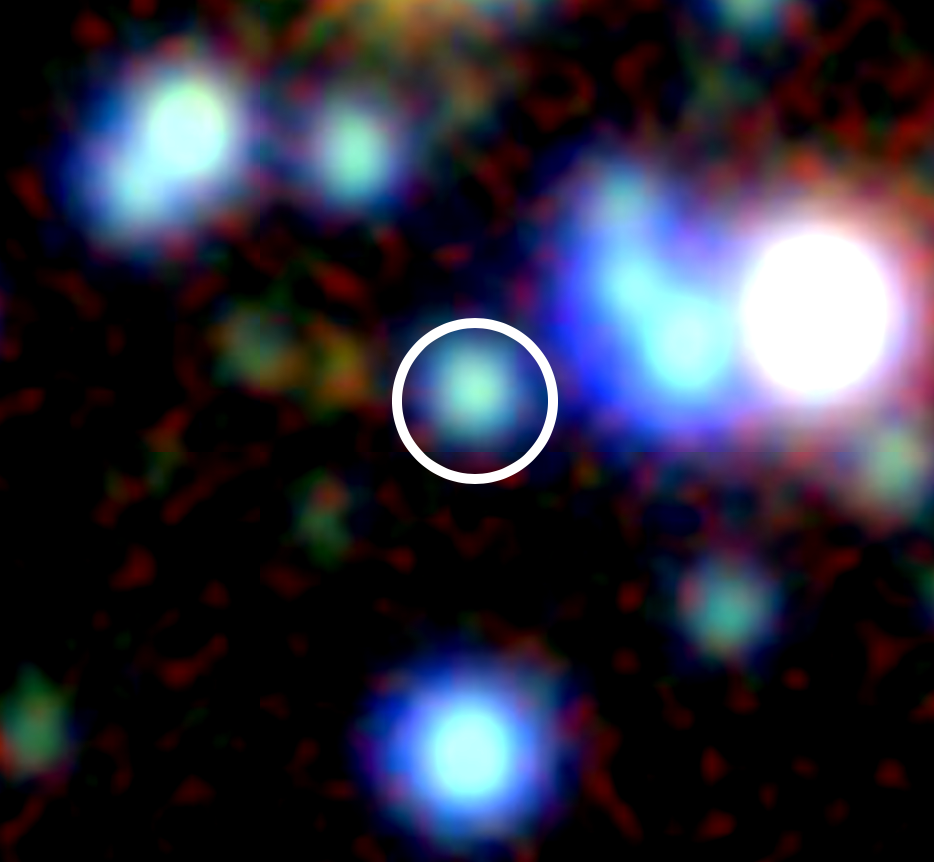}}
  \caption{Optical images around \esrc.
     Top: GROND g' band image. The large white circle (radius 5\arcsec) marks the \ero 
          error circle and the red circle indicates the 1\,$\sigma$  position 
          uncertainty obtained from the \xmm/EPIC X-ray images.
          The small white circle marks the position of the proposed optical counterpart. 
          The three nearest Gaia objects are marked, demonstrating the good astrometric quality
          of the \xmm data.
    Bottom: Zoomed-in archival \decam image centred on the optical counterpart from the top figure. The image is produced from a number of exposures, which were co-added to create 
          the individual filter images for the presented colour composite. 
          The data are taken from the Legacy Survey release 10 (DR10).}
  \label{fig:grond}
\end{figure*}

The total exposures accumulated in the [g, r, i, z] filters are 3.1\,ks, 2.7\,ks, 1.4\,ks, and 0.6\,ks, respectively.
The measured AB magnitudes, an absolute spectral flux density photometric system, are [22.15, 21.82, 21.76, and 21.64] in [g, r, i, and z], respectively, and given in Table\,\ref{tab:grond_mag}. The photometric
calibration is tied to Pan-STARS DR1 \citep{2016arXiv161205560C} through an uber-cal self-calibration method \citep{2012ApJ...756..158S}. The images centred on the position obtained for the counterpart of \esrc are shown in Fig.~\ref{fig:grond}.

\begin{figure}
\centering
  \resizebox{0.9\hsize}{!}{\includegraphics[]{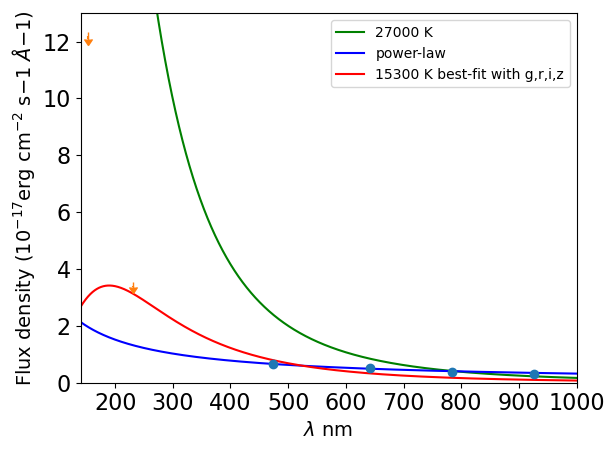}}
  \caption{Optical IR (\decam) flux density of \esrc (blue dots) with the best-fit black-body spectrum in red and a power law overplotted in blue. The UV fluxes from GALEX are plotted as upper limits (in orange) as they were contaminated by a nearby bright object and were not used for the fit. Also overplotted is a black-body spectrum with a temperature of 27000 K in green, as obtained for the case of the DD system RX\, J0806.3+1527 \cite[][]{2014A&A...561A.117E}.}
  \label{fig:SED}
\end{figure}

Figure~\ref{fig:SED} shows the Galactic foreground-corrected SED\footnote{\url{https://www.astronomy.ohio-state.edu/martini.10/usefuldata.html}} 
using the DECam g, r, i, and z-band magnitudes converted to flux densities. GALEX (NUV, FUV) and GROND photometry were not included due to contamination from nearby bright objects. The SED is consistent with a power law of slope of $-$1. Notably, it substantially diverges from the observations of a hot black body associated with the surface of the primary WD for RX\,J0806.3+1527 \cite[][]{2014A&A...561A.117E}. This suggests that the SED of \esrc is either dominated by a different emission region or arises from multiple components. A simple black-body fit is insufficient to describe the data in this case. 

We also examined our \xmm optical/UV monitor (OM) data but could not find a significant source at the optical position of the counterpart. The faintest star that was detected in the OM data has a magnitude of $\sim$18.9 mag. Thus, the non-detection is consistent with the above-derived magnitudes from GROND and archival data.


\begin{table*}
\centering
\caption[]{Optical photometry of the optical counterpart near \esrc.}
\begin{tabular}{lcccccc}
\hline\hline\noalign{\smallskip}
\multicolumn{1}{c}{Name} &
\multicolumn{1}{c}{FUV} &
\multicolumn{1}{c}{NUV} &
\multicolumn{1}{c}{g'} &
\multicolumn{1}{c}{r'} &
\multicolumn{1}{c}{i'} &
\multicolumn{1}{c}{z'} \\
\multicolumn{1}{c}{} &
\multicolumn{1}{c}{(mag)} &
\multicolumn{1}{c}{(mag)} &
\multicolumn{1}{c}{(mag)} &
\multicolumn{1}{c}{(mag)} &
\multicolumn{1}{c}{(mag)} &
\multicolumn{1}{c}{(mag)} \\
\noalign{\smallskip}\hline\noalign{\smallskip}
 
 \decam & -- & -- & 22.2 $\pm$ 0.01 & 21.8 $\pm$ 0.01 & 21.8 $\pm$ 0.02 & 21.6 $\pm$ 0.05 \\
 GALEX & $>$21.5 & $>$22.03 &  &  &  &  \\

\noalign{\smallskip}\hline
\end{tabular}
\tablefoot{GALEX magnitudes are affected by a nearby bright object, and are therefore shown as only upper limits in Fig.~\ref{fig:SED}. Uncertainties correspond to 1\,$\sigma$ level.}
\label{tab:grond_mag}
\end{table*}

\section{Long-term variability}
\label{sec:var}
\begin{figure}
\centering
  \resizebox{1.0\hsize}{!}{\includegraphics[]{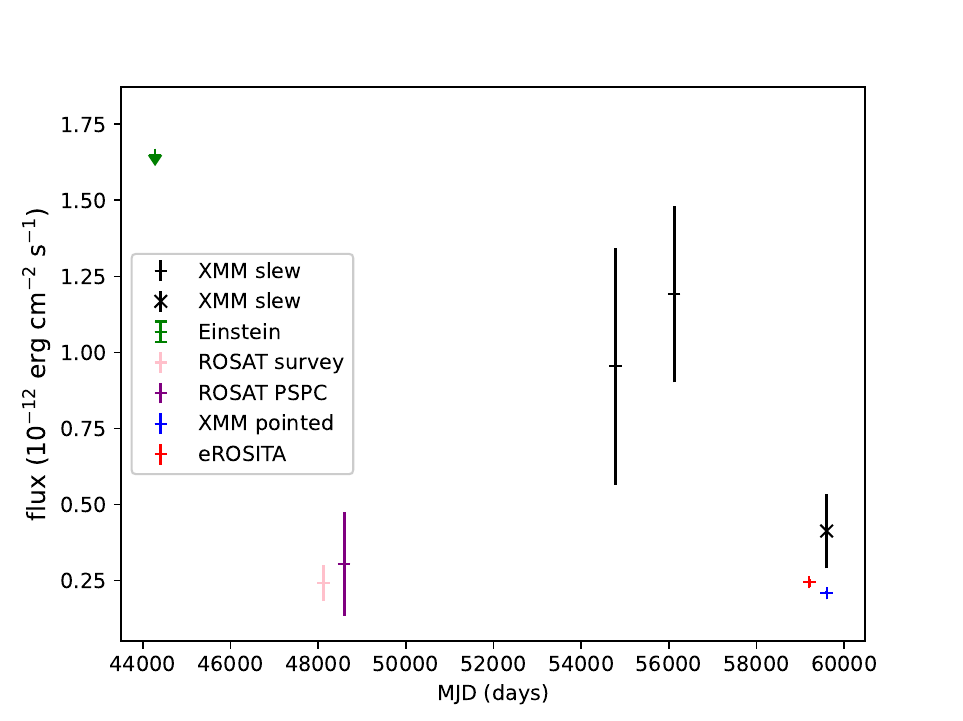}}
  \caption{Long-term X-ray light curve of \esrc for the 0.2–-2\,keV band. The mean value of all observations from eRASS1--4 is marked in red and our \xmm pointed observation in blue. The two sources from the clean XMMSL2 catalogue are marked with a black "+". The black "$\times$" marks the slew observation that belongs to our \xmm pointed observation. A black-body model with a temperature of 100 eV was used for flux conversion for the data from the upper limit server, i.e. for \einstein, \rosat and \xmm slew data. The 1\,$\sigma$ errors are given for \rosat and \xmm slew data, and the \einstein upper limit is given with a 99.7\% confidence level. Errors for \xmm pointed and \ero data have 90\% confidence intervals.}
  \label{fig:long_lc}
\end{figure}

In order to investigate the long-term trend of \esrc, we looked in the HILIGT upper limit server{\footnote{\url{http://xmmuls.esac.esa.int/upperlimitserver}}}.
The position of \esrc was previously covered by the \einstein Observatory, \rosat and 18 \xmm slews. Data for all \xmm slews were downloaded from the XMM-Newton Science Archive{\footnote{\url{http://nxsa.esac.esa.int/nxsa-web/\#search}}} 
and reviewed manually. Significant counts at the source position of \esrc could only be determined for three slews: one belongs to our \xmm pointed observation; the other two are listed in the full and clean \xmm Slew Survey XMMSL2 Source Catalogue{\footnote{\url{https://www.cosmos.esa.int/web/xmm-newton/xsa\#download}}}. Fluxes for all previous observations were taken from the \xmm upper limit server. They were calculated by converting the count rates, assuming a black body with a temperature of 100\,eV and an \nh = 1\hcm{21}. Upper limits are given with a 99.7\% confidence level. The errors are 1\,$\sigma$ errors from the upper limit server. Fluxes and errors for \xmm pointed and \ero data correspond to those determined from the spectral fits (see Table\,\ref{tab:spectral}). The fluxes from the \rosat, \xmm pointed and \ero observations agree within their error margins, while for \xmm slew data marginally higher fluxes were determined. However, the short exposures of the slew observations (5--16 s) likely reflect the short-term variability of the object, as the flux depends strongly on the pulse phase covered by the slew observations. We checked that detections during the slews preferentially occur near pulse maximum (the best-obtained flux at the pulse maximum is a factor of $\sim$3 higher than the phase-averaged flux, see Table\,\ref{tab:spectral}) while slews during phases of low flux lead to non-detections, as many derived upper limits suggest. We also verified this by checking the phase of the slew closest to our \xmm\, observation by using the best-fit periodicity value and confirmed that the times coincide close with the pulse maxima.
Excluding the \xmm slew data, the ratios of the maximum flux from the pointed \rosat and \xmm observations and the \ero data are 1.25$\pm$1.1 and 0.9$\pm$0.6, respectively, consistent with constant flux (Fig.~\ref{fig:long_lc}).

The two sources that are both listed in the clean and full catalogue have distances to \esrc of 4.3\arcsec\ and 6\arcsec\ and positional errors of 4.3\arcsec\ and 3.2\arcsec, respectively. They were recorded with 6.1 $\pm$ 2.5 and 17.9 $\pm$ 4.4 counts in the 0.2--2\,keV band. Both are also detected in the 0.2--12\,keV band with about the same number of counts, but not in the 2--12\,keV band. Therefore, they must be soft sources and could be counterparts of \esrc. However, we cannot fully exclude that the source whose positional error is half its distance to \esrc is a different object.

\section{Discussion}
\label{sec:discussion}

\esrc is a new SSS in the direction of the LMC, discovered through \ero observations. We report an in-depth X-ray and optical study of the source using \xmm, \ero, GROND, and \decam observations. 
We demonstrate that the properties of the source strongly indicate a double degenerate ultra-compact binary system. A periodicity of 374\,s is discovered in the \xmm and \ero data, which we attribute as the orbital period of the system of 6.2\,min.
The system exhibits characteristics similar to the two classical DD systems (see below), RX\,J0806.3$+$1527\,\citep[\CnC,][]{1999A&A...347...47B} and its twin RX\,J1914.4$+$2456\,\citep[\Vul,][]{1995A&A...297L..37H} and the more recently discovered \xsrc in the foreground of the LMC \citep[][]{2017A&A...598A..69H}.

The X-ray spectrum of \esrc can be described with absorbed black-body radiation. The derived temperature of kT $\sim$ 110\,eV is at the slightly higher end but consistent with soft emission seen from SSSs \citep[typical  kT $\sim$ 40--90\,eV,][]{1995A&A...297L..37H,1996A&A...310L..25B}. 
The derived  \nh is lower than the total Galactic value in the source direction (see Table\,\ref{tab:spectral}), suggesting that \esrc is an SSS detected in the foreground of the LMC.
A high density of known SSS in the direction of the Magellanic Clouds can likely be explained by the relatively low Galactic column density and the large number of existing X-ray observations sensitive to low energies \citep{2022A&A...657A..26M}, which has an especially full coverage now with \srg/\ero. 

The X-ray luminosity of the object is too low for it to be classified as a nuclear burning white dwarf \citep[][]{1992A&A...262...97V}.
The presence of a periodicity of 374\,s, the measured X-ray luminosity, and the detection of an additional hard bremsstrahlung spectral component in the \xmm data (Table\,\ref{tab:spectral}) could indicate a cataclysmic variable (CV) nature of the object, in particular an intermediate polar (IP). In IPs, the white dwarf orbits a late-type dwarf. The radiation originates as the accreted plasma couples to the field lines and forms a shock wave close to the WD magnetic poles. The X-ray emission originates from the heated polar caps emitting soft X-rays and the shocked plasma emitting thermal bremsstrahlung.
A periodic signal of 374\,s indicates the spin period of the WD in this case, and is consistent with values found in the literature \citep[e.g.][]{2020ApJ...897...70P}. However, there are several arguments to disfavour an IP nature of \esrc. Firstly, CVs and IPs have bright optical counterparts and display $\log(f_{\rm x}/f_{\rm opt}$) $\lesssim1.5$ \citep[][]{2003A&A...404..301R,2017PASP..129f2001M}. Fainter optical counterparts are usually detected in the case of DD ultra-compact binaries, (see Table\,\ref{tab:dd} to compare the observed $\log(f_{\rm x}/f_{\rm opt}$) values). 

Secondly, the X-ray flux of \esrc drops to zero between pulses, with a duty cycle of $\sim50\%$. This would require an extreme geometric configuration for the IP with an inclination approaching 90$^{\circ}$ and in addition a very high \nh at the off-cycle phases to absorb the flux completely.

Instead, if \esrc is a “polar”, in which the strong magnetic field of the white dwarf locks its spin to the orbit of its companion star, the 374\,s (6.2\,min) periodicity corresponds to the orbital period of the binary system. The above implies that \esrc has one of the shortest periods known for any binary system, only after the 5.4\,min orbital period  of \CnC. Such a small value implies that the donor is a degenerate star, making \esrc a probable member of the AM CVn stars that host double degenerate systems. 
It is most likely that \esrc belongs to this rare class of AM CVns when in the “direct impact” accretion scenario they pass through a short-living ultra-compact phase (DD) when binary periods close to 2--3\,min are possible \citep[][]{2001A&A...368..939N}. Alternatively, in the “uni-polar” inductor model, this phase could last longer \citep[up to $10^{5}$\,yr;][]{2007A&A...464..417D}), depending on the asynchronisation between the spin and the orbital period of the system.
Moreover, the detection of a 100\% modulated pulsed flux and the absence of flux for 50\% of the cycle further testifies in favour of its nature as an ultra-compact DD. The 100\% modulation of the periodic signal can be explained by self–occultation of the stream impact point on the surface of the accreting white dwarf. The long term variation of the mean $L_{\rm x}$ as demonstrated in Sect.~\ref{sec:var} is also consistent with a DD's nature \citep[][and references therein]{2002A&A...386L..13I,2008ApJ...683..375D}.
 
Currently, there are only three other X-ray sources that fall into the category of a DD: namely, \CnC, \Vul, and \xsrc. Due to the compact orbit of the systems, accretion discs are not formed (as the minimum distance from the centre of the donor is smaller than the size of the accretor). A direct impact phase occurs instead, which lasts for a few million years \citep[see][for a review]{2010PASP..122.1133S}. The recent models involving direct impact accretion all require \esrc to have a luminosity greater than \oergs{33}, more likely of the order of \oergs{34} \cite[see e.g. Fig. 4 of][and references therein]{2012ApJ...758...64K}. This requires the system to be at a distance $\gtrsim5$\,kpc. The L$_{\rm x}$ for \esrc corresponds to $\sim$ 6\ergs{31} -- 1.5\ergs{33} (d = 1--5 kpc, see also Table\,\ref{tab:spectral}). This distance range is consistent with the population of old stars detected in the Milky Way's stellar halo \citep[][]{2008ApJ...680..295B}.

 The alternative model to explain the emission from DD is the “unipolar inductor” (UI) model \citep[e.g.][] {2006A&A...447..785D,Colpi2009PhysicsOR} where the system is powered by electric currents (generated between a magnetic primary WD and a non-magnetic secondary that does not fill its Roche lobe). The maximum possible luminosity of \esrc in the case of the UI model is given by \citet[][]{2012ApJ...757L...3L} as 
 \begin{equation}
\label{eq:Lai}
L_{\rm max} \approx 10^{32} \zeta_{\phi} \left(\frac{\Delta \Omega}
{\Omega}\right) \mu^2_{32} R^2_9 \left(\frac{{M}_{\mathrm{tot}}}{{M}_{\odot}}\right)^{-5/3} \left(\frac{{P}}{6~{\rm
    min}}\right)^{-13/3}~\mathrm{erg~s}^{-1},
\end{equation}
where $\zeta_{\phi}$ depends on the degree of bending of the flux tube ($\zeta_{\phi}=1$ for a maximally bent tube), $\Delta \Omega$ is the difference between the orbital frequency ($\Omega$) and the primary spin frequency, $\mu$ is the magnetic moment of the primary, $R$ is the secondary radius, $M_{\rm tot}$ the total mass of the system, and $Q_X$ stands for a quantity, $Q$, in units of $10^X$. The estimate $L_{\rm x}$ of \esrc is consistent with the expectations of the UI model for a distance between 1--2\,kpc, and can explain the X-ray emission from the object. 

Further observational evidence like deep optical observations to detect the orbital period and determine its phase with respect to X-rays are required to distinguish between the direct impact and UI models; for example, a varying phase offset detected between the optical and X-ray modulation can be naturally explained by the direct impact model \citep[see discussion in][]{2014A&A...561A.117E}. Determination of the orbital period evolution and a robust estimate of the period derivative is also crucial to constrain the models for DD \citep[][and references]{2004MNRAS.350..113M}. Another key to understanding the physical nature of DD is the possible detection of a hard bremsstrahlung component in the spectrum, especially in the off-pulse phase (see Fig.~\ref{fig:pp} and Table\,\ref{tab:spectral}), which is one of the first such detections for this class of objects. The detection of a hard X-ray component could pave the way to further refining the direct impact shock regions, 
white dwarf masses in the system, and the specific accretion rates \citep[][]{2008ApJ...683..375D}.

\begin{table*}
\centering
\caption[]{Properties of DD ultra-compact binaries}
\begin{tabular}{llllllllc}
\hline\hline\noalign{\smallskip}
\multicolumn{1}{l}{Source} &
\multicolumn{1}{l}{Orbital Period [min]} &
\multicolumn{1}{c}{kT [eV]} &
\multicolumn{1}{c}{L$_{\rm x}$ [\uergs]} &
\multicolumn{1}{c}{$\log(f_{\rm x}/f_{\rm opt}$)} &
\multicolumn{1}{c}{g'-r'} &
\multicolumn{1}{c}{References}\\
\noalign{\smallskip}\hline\noalign{\smallskip}

\CnC      & 5.4           & 65        & $^{a}$4.5$\times10^{34}$   &$^{e}$2.3 (2.4) & -0.44 & 1,2,7 \\
 \Vul       & 9.5           & 43        &  $^{b}$5$\times10^{35}$- 4$\times10^{36}$  & $^{e}$1.6 (3.5) & 0.99 & 3,4,7,8   \\
 \xsrc                  &   23.6         & 69           & $^{a}$5$\times10^{32}$     & $^{e}$1.1 (1.2)  & -0.03 & 5,6 \\
 \esrc &   6.2             & $^{d}$110        & $^{c}$6$\times10^{31}$-$1.5\times10^{33}$ &$^{e}$1.6 (1.9) & 0.25  & this work \\
\noalign{\smallskip}\hline
\end{tabular}
\tablefoot{References:
 
1) \citet{2003ApJ...598..492I};
2) \citet{2005A&A...433..635S};
3) \citet{1995A&A...297L..37H};
4) \citet{1998MNRAS.293L..57C};
5) \citet{2018A&A...617A..88R};
6) \citet{2017A&A...598A..69H};
7) \citet{2007MNRAS.374.1334B};
8) \citet{2008MNRAS.384..687R};

} 
\tablefoot{$^{a}$: for d=5\,kpc; $^{b}$: for d=4--5\,kpc, \gaia geometric distance 4.8$^{+1.8}_{-1.6}$ kpc \citep{2021AJ....161..147B}; $^{c}$: for d=1--5\,kpc; $^{d}$: also a hard bremmstahlung component detected; $^{e}$: computed with the information of the mean g' magnitude and maximum observed $F_{\rm x}$ (un-absorbed and reddening-corrected). The g'-r' colours were obtained after reddening correction. The luminosities were corrected for absorption.}
\label{tab:dd}
\end{table*}

Table\,\ref{tab:dd} summarises the X-ray and optical properties of \esrc as compared to the three classical DDs \CnC, \Vul, and \xsrc. Although their distances are not well known, it seems that \Vul is significantly more luminous in X-rays than the other three objects. The bolometric luminosity is however uncertain, given the high \nh in \Vul.
 The X-ray spectra of all three systems can be described by an absorbed black-body model; however, the measured black-body temperature and the detection of a hard tail in \esrc\ indicates that it is the hottest object of the three.
It is also noteworthy that the pulse profiles of \esrc closely resemble those of \CnC and \Vul, rising steeply to the maximum, declining by half over the width of the pulse, and then dropping steeply again to zero. In the case of \xsrc, on the other hand, which has a slightly longer orbital period of 23.6 min, the profile is reversed, with a fast decay and a slower rise.
Modelling the morphology of the pulse profiles can provide clues about the location of the emission region, and the size and extent of the X-ray emitting hotspot \citep[see discussions in][]{2018A&A...617A..88R}.

Finally, the identification of \esrc as a DD ultra-compact binary provides valuable information on the rapidly growing class of this rare evolutionary phase in AM CVns. Understanding the evolution of AM CVns is also important as a fraction of them may create thermonuclear supernovae on timescales of $\sim$10$^{8}$\,yr, either as .Ia or Ia \citep[][]{2014MNRAS.438L..26K,2018MNRAS.473.5352L}. Furthermore, as systems with the shortest orbital periods ever recorded, DD ultra-compact binaries represent one of the most promising targets for persistent gravitational wave detection \citep[][]{2001A&A...368..939N} and will be used in the verification phase of space-based gravitational wave observatories such as LISA \citep[][]{2017MNRAS.470.1894K,2020ApJ...893....2L}.

\begin{acknowledgements}
We thank the referee for useful comments and suggestions.
This work is based on data from \ero, the soft X-ray instrument aboard \srg, a joint Russian-German science mission supported by the Russian Space Agency (Roskosmos), in the interests of the Russian Academy of Sciences represented by its Space Research Institute (IKI), and the Deutsches Zentrum f{\"u}r Luft- und Raumfahrt (DLR). The \srg spacecraft was built by Lavochkin Association (NPOL) and its subcontractors, and is operated by NPOL with support from the Max Planck Institute for Extraterrestrial Physics (MPE).
The development and construction of the \ero X-ray instrument was led by MPE, with contributions from the Dr. Karl Remeis Observatory Bamberg \& ECAP (FAU Erlangen-N{\"u}rnberg), the University of Hamburg Observatory, the Leibniz Institute for Astrophysics Potsdam (AIP), and the Institute for Astronomy and Astrophysics of the University of T{\"u}bingen, with the support of DLR and the Max Planck Society. The Argelander Institute for Astronomy of the University of Bonn and the Ludwig Maximilians Universit{\"a}t Munich also participated in the science preparation for \ero.
The \ero data shown here were processed using the \eSASS software system developed by the German \ero consortium. 
This work used observations obtained with \xmm, an ESA science mission with instruments and contributions directly funded by ESA Member States and NASA. The \xmm project is supported by the DLR and the Max Planck Society.
This research has made use of data obtained from XMMSL2, the Second \xmm Slew Survey Catalogue, produced by members of the XMM SOC, the EPIC consortium, and using work carried out in the context of the EXTraS project ("Exploring the X-ray Transient and variable Sky", funded from the EU's Seventh Framework Programme under grant agreement no. 607452). 
Work on \nicer science at NRL is funded by NASA and the Office of Naval Research.
Part of the funding for GROND (both hardware as well as personnel) was generously granted from the Leibniz-Prize to Prof. G. Hasinger (DFG grant HA 1850/28-1).
GV acknowledges support by H.F.R.I. through the project ASTRAPE (Project ID 7802).
This research has made use of the VizieR catalogue access tool, CDS, Strasbourg, France (DOI: 10.26093/cds/vizier). 
The original description of the VizieR service was published in \citep{2000A&AS..143...23O}.

\end{acknowledgements}

\bibliographystyle{aa} 
\bibliography{references} 

\end{document}